\documentstyle[prb,aps,amssymb,psfig]{revtex}
\begin{document}
\draft
\title{Cohesion and conductance of disordered metallic point contacts} 
\author{J.~B\"urki}
\address{Institut de Physique Th\'{e}orique, Universit\'{e} de Fribourg,
CH-1700 Fribourg, Switzerland and \\
Institut Romand de Recherche Num\'erique
en Physique des Mat\'eriaux, CH-1015 Lausanne, Switzerland}
\author{C.~A.~Stafford}
\address{Physics Department, University of Arizona, 1118 E.\ 4th Street,
Tucson, AZ 85721}
\author{X.~Zotos}
\address{Institut Romand de Recherche Num\'erique en Physique des
Mat\'eriaux,
CH-1015 Lausanne, Switzerland}
\author{D.~Baeriswyl}
\address{Institut de Physique Th\'{e}orique, Universit\'{e} de Fribourg,
CH-1700 Fribourg, Switzerland}
\date{Submitted to Phys. Rev. B, 25 February 1999, resubmitted 4 May 1999}
\twocolumn[\hsize\textwidth\columnwidth\hsize\csname@twocolumnfalse\endcsname
\maketitle
\begin{abstract}
The cohesion and conductance of a point contact in a two-dimensional metallic
nanowire are investigated in an
independent-electron model with hard-wall boundary conditions.
All properties of the nanowire are related to the
Green's function of the electronic scattering problem, which is
solved exactly via a modified recursive Green's function algorithm.
Our results confirm the validity of a previous approach based
on the WKB approximation for a long constriction, but find an
enhancement of cohesion for shorter constrictions.
Surprisingly, the cohesion persists even after the last conductance
channel has been closed. For disordered nanowires, a statistical
analysis yields well-defined peaks in the conductance histograms
even when individual conductance traces do not show well-defined
plateaus. The shifts of the peaks below integer multiples of
$2e^2/h$, as well as the peak heights and widths, are found 
to be in excellent agreement with predictions based on random matrix theory,
and are similar to those observed experimentally. 
Thus abrupt changes in the wire geometry are not necessary for
reproducing the observed conductance histograms.
The effect of disorder on cohesion is found to be quite
strong and very sensitive to the particular configuration of impurities 
at the center of the constriction. 
\end{abstract}

\pacs{PACS numbers: 73.23.-b, 73.20.Dx, 72.80.Ng, 73.40.Jn}

\vskip2pc]


\section{Introduction}\label{sec:intro}

In a pioneering experiment published in 1996, Rubio, Agra\"{\i}t
and Vieira \cite{rubio96} simultaneously measured the force and
conductance during deformation and rupture of gold nanocontacts.
They observed nano-Newton force oscillations correlated with
conductance steps of order $2e^2/h$. Similar results were
obtained independently by Stalder and D\"urig.\cite{stalder96}
An explanation of these force fluctuations, based on the response
of the conduction electrons to the mechanical deformation of the
contact, was proposed by Stafford et al. \cite{stafford97a}
In this paper, we give an exact numerical solution to the model
of Ref.\ \onlinecite{stafford97a}, allowing the treatment of any
shape of the constriction and the inclusion of disorder.

Conductance quantization has been widely studied in the past decade,
both experimentally
\cite{wees88,wharam88,olesen94,krans95,brandbyge95,costa-kramer97a,costa-kramer97b,costa-kramer97c}
and theoretically.
\cite{brandbyge95,glazman88,szafer89,he89,buttiker90,beenakker91,maslov93,todorov93,torres94,beenakker94,garcia-mochales97,brandbyge97a,bascones98a}
On the experimental side,
it was first measured in a two-dimensional
electron gas split-gate and was understood to be a consequence of the
quantization of the transverse motion.\cite{wees88,wharam88}
Each transverse energy subband defines a conduction
channel that contributes $2e^2/h$ to the conductance.
More recently, conductance quantization in units of  $2e^2/h$
has been observed in three-dimensional metallic quantum wires
using various experimental techniques.
\cite{olesen94,krans95,brandbyge95,costa-kramer97a,costa-kramer97b,costa-kramer97c}
In this case, the quantization is not so clear-cut,
and measurements are in general less reproducible.
Therefore, a statistical analysis of a large set of
experimental data is usually made.
The resulting conductance histograms show  well-defined peaks
with centers shifted below integer multiples of $2e^2/h$.
The shifts can be corrected by subtracting a series resistance
of a few hundred Ohms.
This resistance was originally attributed to the bulk contacts,
which are not part of the nanocontact and should thus be subtracted.
\cite{krans95,costa-kramer97a,costa-kramer97b,costa-kramer97c}
It has recently been argued, based on theoretical work,
\cite{garcia-mochales97} that this resistance could also be caused by
disorder in the nanowire, in agreement with previous
studies of quantum point contacts.\cite{maslov93,beenakker94}

On the theoretical side, the problem of conductance quantization has
been investigated using a variety of techniques, both in two and
three dimensions.
\cite{brandbyge95,glazman88,szafer89,he89,buttiker90,beenakker91,maslov93,todorov93,torres94,beenakker94,garcia-mochales97,brandbyge97a,bascones98a}
It is found that conductance plateaus are well defined only in
constrictions that are smooth on the scale of the Fermi wavelength.
\cite{glazman88,buttiker90,torres94}
Furthermore, the plateaus can
also be affected by thermal fluctuations or disorder.
Thermal effects, which smoothen out the conductance steps,\cite{he89}
are important in the case of a two-dimensional electron gas,
\cite{wees88,wharam88} where the spacing between subbands is of the
order of $1 K$, but negligible in the case of metallic point contacts,
\cite{rubio96,stalder96,olesen94,krans95,brandbyge95,costa-kramer97a,costa-kramer97b,costa-kramer97c}
where the spacing between subbands is of the order of $10^4 K$.
The effect of disorder on conductance quantization has been studied both
analytically\cite{maslov93,beenakker94,bascones98a}
and numerically.\cite{brandbyge95,garcia-mochales97,brandbyge97a}
Disorder was found to reduce the conductance compared to the ballistic case,
leading to shifts of the peaks in the conductance histograms similar
to those observed experimentally.
We note that most of the theoretical
conductance curves were obtained by changing the Fermi energy,
in contrast to the experimental situation for a metallic nanowire,
where the Fermi energy is an invariant property of the material and the
shape of the contact is modified.

The structural transformations of  metallic nanocontacts were
first studied by Landman and coworkers,\cite{landman90} and later
by others,\cite{brandbyge95,todorov93} using molecular dynamics.
These simulations suggest that the elongation of a connective neck
proceeds through a sequence of abrupt structural transformations,
involving a succession of elastic and yielding stages. Within this approach,
the conductance is expected to change abruptly due to
atomic rearrangements,\cite{brandbyge95,todorov93} and the force
is expected to follow a sawtooth behavior.\cite{brandbyge95,landman90}
The molecular dynamics simulations give rise to the following
interpretation\cite{rubio96,todorov96,landman96a} of experiments:
During elastic stages, the conductance is essentially constant
at a quantized value and the magnitude of the force increases with a 
constant slope,
while the yielding stages consist of abrupt relaxations of the
atomic structure, giving rise to a sudden change of conductance and
force. It was asserted that both the elastic and transport properties
of narrow metallic constrictions can be understood on the basis
of atomic rearrangements, calculated in the framework of
classical lattice dynamics.

It thus came as a surprise when it was discovered\cite{stafford97a}
that a simple free electron model, which neglects the discrete atomic
structure of the constriction, is able to reproduce both the
oscillations in the tensile force and the conductance plateaus
of Ref.\ \onlinecite{rubio96}. In this model, each conductance
channel is viewed as a long chemical bond which is stretched and
broken, giving rise to the observed force oscillations.
Similar results were obtained by van Ruitenbeek et al.\cite{ruitenbeek97b} 
using a free-electron model
and by Yannouleas and Landman,\cite{yannouleas97} using a slightly
more realistic jellium model. However, these studies dealt with
closed systems, while the experimental nanowire is a contact between
two macroscopic pieces of metal, and is thus an open system. Since
mesoscopic effects can be very different in the canonical and
grand canonical ensembles, this is an important distinction.

In this paper, we give an exact numerical solution of the
free-electron model proposed in Ref.\ \onlinecite{stafford97a}.
Our formalism allows us to extend the analytical results of
Ref.\ \onlinecite{stafford97a} by treating both non-adiabatic
constrictions and disordered wires. We find that the cohesion of
short (non-adiabatic) constrictions is increased compared to that
of long constrictions. For disordered systems, we obtain conductance
histograms which look very similar to those observed experimentally.
Our results also suggest that the cohesion of nanowires is
quite sensitive to disorder.

The rest of this paper is organized as follows: In section
\ref{sec:model}, we describe our model in which conductance and cohesion
are interconnected through the scattering matrix.
Section \ref{sec:method} presents the numerical method,
while the results are discussed in section \ref{sec:results}.
A short summary is given in section \ref{sec:conclusions}.

\section{Model and basic quantities}\label{sec:model}

This section introduces our free-electron model. Its simplicity
allows to treat both the conductance and the metallic cohesion
in a unified way.

\subsection{Assumptions}\label{subsec:asssumptions}

Our model is based on several simplifying assumptions.
{\sl (i) Free electrons:}
Gold is known to be rather well described by the model of free electrons
where the atomic potentials are neglected.\cite{ashcroft76}
This indicates that the detailed configuration of atoms has little
effect on both electronic and cohesive properties. By adopting the
free electron model, we eliminate the kinetics of atomic
rearrangements, which seem to play a role in actual experiments,
at least for thick constrictions.\cite{agrait95}
{\sl (ii) Independent electrons:}
We neglect electron-electron interactions. Recent calculations,
one based on an extended Thomas-Fermi scheme,\cite{yannouleas98}
the other on the Hartree approximation, \cite{stafford98} suggest
that interactions have little effect on the quantities considered
here. However, more careful studies will be needed to substantiate
this conclusion, as interaction effects are particularly subtle in
one-dimensional systems and thin wires.
{\sl (iii) Smooth constriction:}
According to molecular dynamics simulations
\cite{brandbyge95,todorov93,landman90,landman96a} and experiments,
\cite{kondo97,kizuka98} the shape of elongated gold nanowires is
quite regular. Therefore we model the contact as a smooth
constriction, which changes continuously upon deformation.
{\sl (iv) Constant volume:}
The volume of the constricted part is assumed to be conserved
during the deformation, in agreement with molecular dynamics
simulations.\cite{landman90} Recent experiments\cite{yanson98}
indicate that this assumption is not always valid, for instance
in the case of mono-atomic chains.
{\sl (v) Hard-wall boundary conditions:}
Electrons are confined to a wire due to a smooth attractive ionic
potential. We approximate the potential in terms of hard-wall boundary
conditions, keeping in mind that the radius of the boundary will be
larger than the effective radius defined by the electronic density.
\cite{ruitenbeek97b}
{\sl (vi) Zero temperature:}
The thermal population of electronic subbands above the Fermi energy is
negligible for nanowires. Therefore we consider the zero-temperature
limit, where inelastic scattering processes, e.g. due to phonons or
electron-electron collisions, are absent.

\subsection{Model}\label{subsec:model}

The physics of the problem is similar for two- and three-dimensional
wires except that, for a two-dimensional system, {\sl (i)} the
transverse energy levels are nondegenerate, and therefore the steps
in the conductance curves are all of height $2e^2/h$, and {\sl (ii)}
the surface energy is reduced as compared to that of a
three-dimensional wire; this changes the overall slope of the force
curves. We restrict ourselves to two dimensions, where the
computational effort is much lower.

\begin{figure}
\begin{center}
\psfig{figure=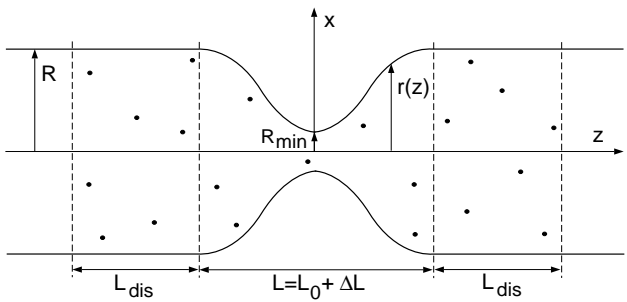,width=8.6cm}
\end{center}
\vspace{-0.5cm}
\caption{
Schematic diagram of a constriction in a 2D quantum wire.
Electrons are confined along the $z$ axis by a hard-wall potential
at $x=\pm r(z)$.  Localized impurities of potential $V(x-x_j,z-z_j)$
are distributed randomly in the constriction and within a part
of length $L_{dis}$ on each side of it.}
\label{fig:geometry}
\end{figure}

Our model wire is described by the eigenvalue equation
\begin{equation}
\left[-\frac{\hbar^2}{2m}\left(\partial_x^2+\partial_z^2\right)
+V_{dis}(x,z)\right]\psi(x,z)=E\psi(x,z),
\label{eq:H}
\end{equation}
where $x$ and $z$ are the transverse and longitudinal coordinates,
respectively, and $V_{dis}(x,z)$ is a potential due to disorder.
We use the boundary condition $\psi=0$ for $x^2 = r^2(z)$,
where $r(z)$ defines the geometry (see Fig.\ \ref{fig:geometry}). 
In the initial configuration,
there is no constriction ($r(z)=R$) and the disorder is restricted
to a finite part of the wire (namely to $|z|< (L_0/2+L_{dis})$,
where $L_0$ is the initial length of the deformable part).
After elongation by $\Delta L=L-L_0$,
the shape of the deformable part is chosen to be, for $|z|<L/2$,
\begin{equation}
r(z)=R_{min}+(R-R_{min})\left(3u^2-3u^4+u^6\right),
\label{eq:rz}
\end{equation}
where $u=\left|\frac{2z}{L}\right|$
and $2 R_{min}$ is the width of the narrowest part of the
constriction, related to $L$ by the condition of constant volume.
The specific choice of the
function $r(z)$ is not crucial,\cite{stafford97a} as has been checked by 
using other shapes. 

The potential $V_{dis}(x,z)=\sum_j V(x-x_j,z-z_j)$ describes
randomly distributed impurities, with a given density $n_{i}$.
It can represent structural defects or real impurities. Similar effects
are expected to arise due to surface roughness.\cite{brandbyge95} During
deformation, the impurities are moved in such a way that their local
concentration remains constant. Since we are dealing with an open
system, we use the grand-canonical ensemble and take the  chemical
potential to be the Fermi energy of the macroscopic wires connected
to the nanocontact. Throughout this paper, we consider the case of
gold, with a Fermi energy $\varepsilon_F=5.5\,eV$.

\subsection{$S$ matrix}\label{subsection:formalism}

The formalism developed by Stafford et al. \cite{stafford97a}
allows one to treat both transport and mechanical properties of open
mesoscopic systems on an equal footing, in terms of the electronic
scattering matrix $S$. This formalism is general and can be applied
to any non-interacting open system connected to an arbitrary number
of leads.

The conductance for the present two lead system
is calculated using the Landauer formula\cite{fisher81}
\begin{equation}
\label{eq:landauer}
G=\frac{2e^2}{h}\text{Tr }(t^{\dagger}t),
\end{equation}
where $t$ is the transmission matrix at the Fermi energy
(directly related to the $S$ matrix).

The force is obtained from the variation of the appropriate free energy, the
grand canonical potential
$\Omega(\epsilon_F,L)$, with respect to the length $L$ of the
deformed region. For a non-interacting system of electrons, one has
\begin{equation}
\label{eq:energy}
\Omega(\epsilon_F,L) = \int^{\epsilon_F}\!dE \left(E-\epsilon_F\right) D(E),
\end{equation}
where the density of states $D(E)$ is related to the $S$ matrix by
\cite{dashen69,gasparian96}
\begin{equation}
\label{eq:dosgen}
D(E)=\frac{1}{2\pi i}\text{Tr }\left\{S^{\dagger}(E)\frac{\partial S}
{\partial E}-S(E)\frac{\partial S^{\dagger}}{\partial E}\right\}.
\end{equation}
The $S$ matrix provides a fundamental link between conductance and
cohesion, and it is amenable to analytical calculations.\cite{stafford97a} 
$S$ is directly related to the retarded electronic Green's 
function,\cite{dashen69,gasparian96} 
which may be calculated numerically, as explained in the next section.
A closely related formulation of the free energy in terms of the $S$ matrix
was given by Akkermans et al.\cite{akkermans91}

\section{Extended recursive Green's function method}\label{sec:method}

In this section, we describe our numerical procedure. It consists of
computing recursively the Green's function, from which both the
transmission coefficients and the density of states are obtained.

\subsection{Green's function for a wire of constant width}
\label{subsec:method1:cstR}

In order to illustrate our method, we first consider an infinite,
two-dimensional wire of constant width $2R$. In order to proceed numerically,
the original continuum
model is replaced by a discrete lattice, with a width 
of $M$ sites, while the disordered part is restricted to a length of
$N$ sites. This finite region consists of slices, numbered from
$1$ to $N$, each slice representing a cross section with $M$ sites.
The discretized Hamiltonian (for the disordered part of the wire)
is then a ($NM\times NM$) matrix,
with $M\times M$ sub-matrices ${\bf H}_i$ for the $i$-th slice
and $M\times M$ sub-matrices ${\bf V}_{ij}$ for the hopping terms
between neighboring slices $i$ and $j$.

The Recursive Green's Function (RGF) method, developed by MacKinnon,
\cite{mackinnon85} is based on the Dyson equation for the Green's
function. At each step of the iteration, only the relevant parts of the
Green's function are retained. Thus the RGF method allows one to calculate
both the density of states and the transmission coefficients at low
memory cost. Applied to our problem, the method allows us
to construct the Green's function of the disordered part of the wire
slice by slice. The infinite ordered parts are included through
boundary conditions for the first and last slices of the disordered
region (see Ref.\onlinecite{mackinnon85} for details).
The advantage of this method is that one only needs to keep track of a
few $M\times M$ matrices instead of the ($NM\times NM$) matrix
representing the full Green's function.

\subsection{Generalization to a wire of variable width}
\label{subsec:method2:varR}

In order to generalize the RGF method to the case of a wire with
varying thickness, we first apply a scale transformation. Consider
the eigenvalue equation (\ref{eq:H}) together with the boundary
condition $\psi(x,z) = 0$ for $|x| = r(z)$. The change of coordinates
\begin{equation}
\tilde{x}=x\frac{R}{r(z)},\quad \tilde{z}=z,
\label{eq:chgt-var}
\end{equation}
transforms the wire geometry into a stripe of constant width $2R$,
at the cost of a more complicated differential operator,
\begin{eqnarray}
\partial_x^2+\partial_z^2=
\frac{1}{2}\left\{\frac{R^2}{r^2(\tilde z)} + \tilde x^2 \rho^2 ,
\partial^2_{\tilde x}\right\}&+&\partial^2_{\tilde z}
-\left\{\tilde x \rho, \partial_{\tilde x} \partial_{\tilde z} \right\}-
\nonumber \\
&-&\frac{3}{4} \rho^2 +\frac{1}{2}\frac{d\rho}{d\tilde z},
\label{eq:Op}
\end{eqnarray}
where $\rho ({\tilde z}) = \frac{d}{d{\tilde z}}\mbox{log}[r({\tilde z})]$ and
$\{.,.\}$ is the anticommutator. This operator is well-behaved if
$r(z)$ is twice continuously differentiable.

The stripe geometry now allows a straightforward
discretization. For the case of interest, where $r(z) = R$ outside
of the constriction, the RGF method can again be used, as described
above. The transformation (\ref{eq:Op}) has the additional advantage
that the narrowest part of the wire, where the main effects considered
here originate,\cite{stafford97a} is scanned with the finest grid.

\subsection{Calculation of conductance and force}
\label{subsec:method3:cond+force}

Once the retarded Green's function  ${\bf G}_{ij}$ is determined, the
transmission matrix in the Landauer-B\"uttiker formula
[Eq.\ (\ref{eq:landauer})] can be easily calculated
(see Ref.\ \onlinecite{liu94}). The resulting equation for the
conductance is found to be
\begin{equation}
G(L) = \frac{2e^2}{h}\sum_{\mu,\nu}
\left\{\left|\frac{\partial E_{\mu}}{\partial k}
\cdot\frac{\partial E_{\nu}}{\partial k}\right|
|({\bf G}_{0L})_{\mu \nu}|^2\right\}_{E=\varepsilon_F},
\label{eq:conductance}
\end{equation}
where $E_{\nu}(k)$ is the dispersion relation for the $\nu$-th
subband. The Green's function also yields directly the density of states
\begin{equation}
D(E)=-\frac{2}{\pi}\mbox{Im}\mbox{Tr}\, G(E),
\end{equation}
from which the grand canonical potential 
[Eq.\ (\ref{eq:energy})] is obtained by
integration. The tensile force is then calculated by differentiating
the energy $\Omega$ with respect to the length $L$ of the constriction,
\begin{equation}
F(L) = -\frac{\partial\Omega(L)}{\partial L} .
\label{eq:force}
\end{equation}
For the conductance, it is sufficient to know the Green's function
at the Fermi energy. The evaluation of the force requires much more
work, as all occupied electronic states
contribute to the grand canonical potential [Eq.\ (\ref{eq:energy})].
It is thus necessary to recalculate the Green's function
many times, and this is the part that uses most of the computational
effort. Fortunately, the energy $\Omega(L)$ turns out to be a smooth
function of $L$, so that the numerical differentiation can be carried
out for rather large steps $\Delta L$.

\section{Results}\label{sec:results}

In Ref.\ \onlinecite{stafford97a}, we solved the
free-electron model without disorder analytically, using both the adiabatic and
WKB approximations. In this section, the validity of our analytical
results is confirmed numerically for long and clean constrictions.
We then proceed to the case of short necks, where the adiabatic
approximation breaks down. Subsequently, impurities are introduced
in terms of randomly distributed short-ranged potentials. Both
non-adiabaticity and disorder are found to produce interesting new 
effects.

\subsection{Comparison with analytical results}\label{subsec:res-adiab}

We first want to check the analytical results of
Ref.\ \onlinecite{stafford97a}. We thus consider a clean wire
with a long constriction, i.e., with a small enough ratio $R/L_0$,
where $2 R$ and $L_0$ are, respectively, the width of the wire and
the initial length of the constriction. The width is fixed to $k_FR=6$.
We have varied the lattice constant $a$ of our mesh in order to
study possible discreteness effects. We have found that these effects
are negligible for $k_Fa \leq 0.7$. All results presented below
were obtained for lattice constants satisfying these conditions.

In Ref.\ \onlinecite{stafford97a}, it was shown that
within the adiabatic and WKB approximations the force is
invariant with respect to a stretching of the geometry
$r(z)\rightarrow r(\lambda z)$, and thus
\begin{figure}
\begin{center}
\psfig{figure=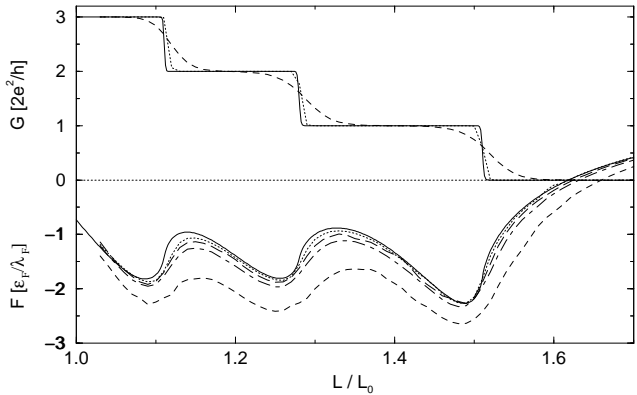,width=8.6cm}
\end{center}
\vspace{-0.5cm}
\caption{Comparison of analytical (WKB) and numerical
(RGF) results for a wire of width $2k_FR = 12$. 
For the conductance, the WKB result ($k_FL_0 = 120$, solid line) 
is compared with two RGF results 
($k_FL_0 =120$, dotted line; $k_FL_0 =12$, dashed line). 
For the force the WKB result (solid line, independent of $L_0$) 
is displayed together with several RGF results 
($k_FL_0 = 120,36,24,12$, from top to bottom).}
\label{fig:adiab}
\end{figure}
\vspace*{-0.7cm}
\begin{figure}
\begin{center}
\begin{tabular}{cc}
\psfig{figure=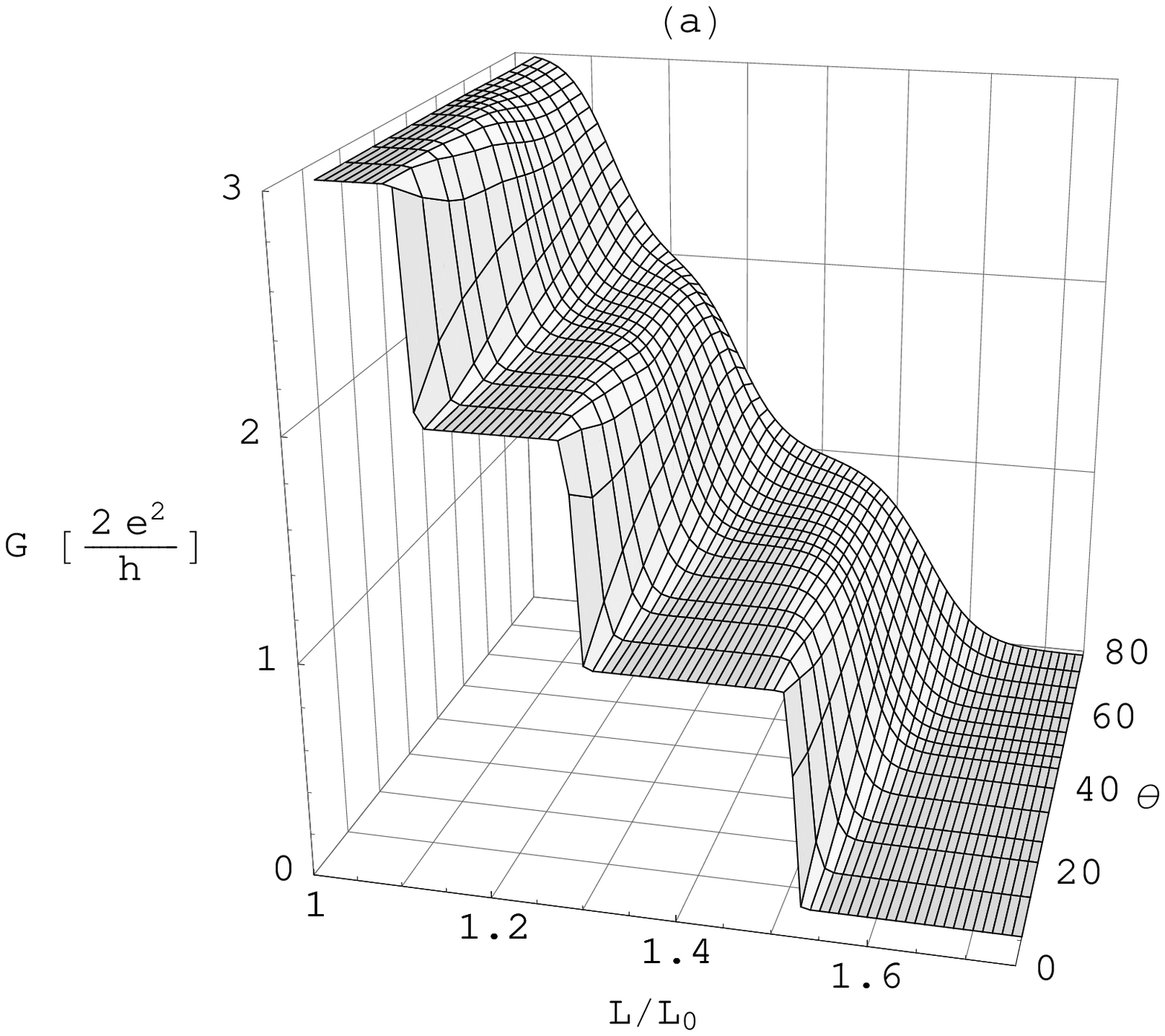,width=8.6cm} &
%
\psfig{figure=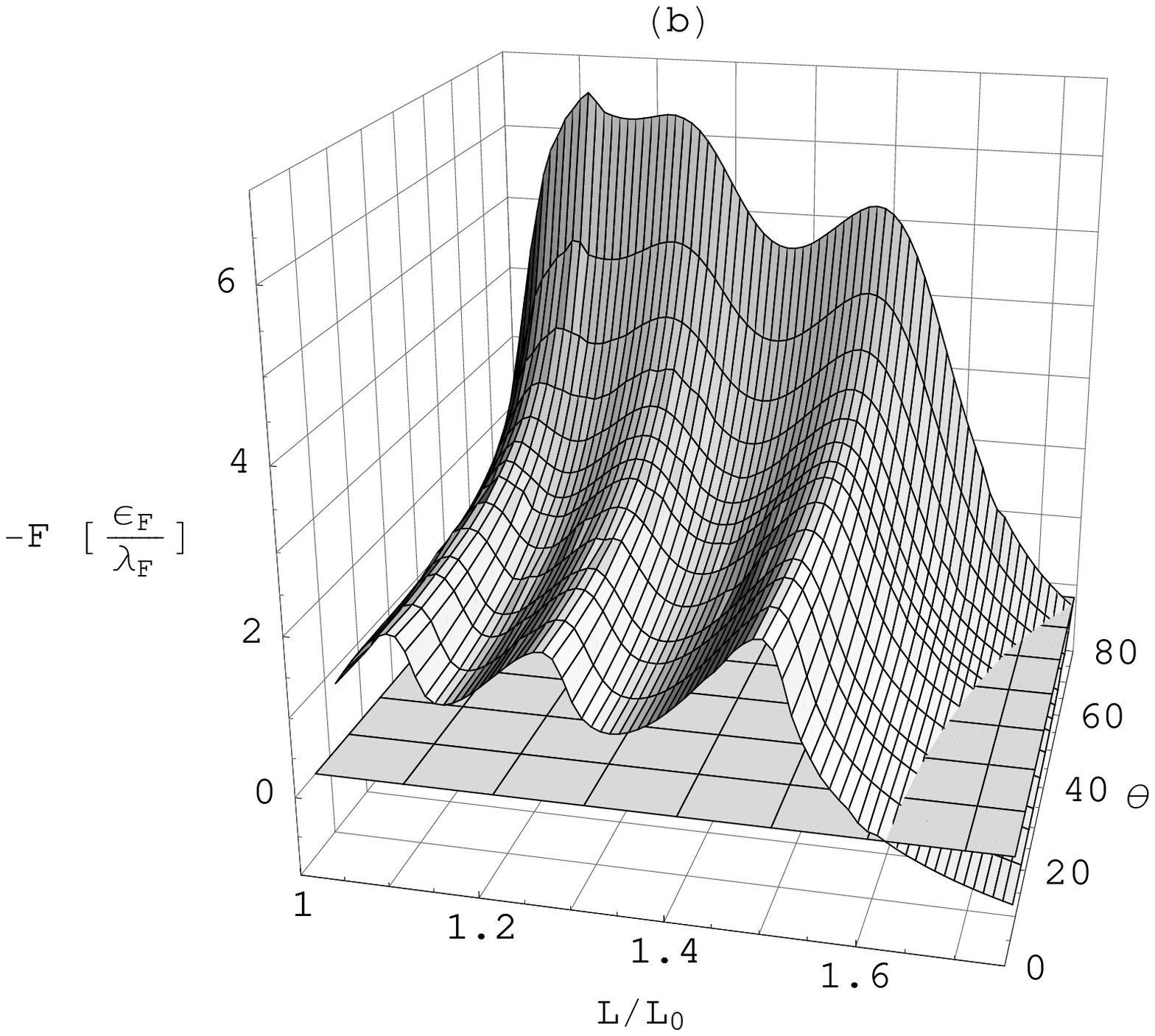,width=8.6cm}
\end{tabular}
\end{center}
\widetext
\caption{Conductance (a) in units of $2e^2/h$
and force (b) in units of $\varepsilon_F/\lambda_F$ as functions of
elongation $L/L_0$ and opening angle $\theta$ for 
a fixed width $2k_F R = 12$.}
\narrowtext
\label{fig:nonadiab}
\end{figure}
%
\noindent is independent of the initial length $L_0$.
We therefore compare in Fig. \ref{fig:adiab}
the analytical results for a single initial length, $k_FL_0 = 120$,
with numerical results for different $L_0$.
For $k_FL_0 = 120$ the conductance curves are almost indistinguishable,
and the numerical result for the force comes very close to the
analytically calculated curve. The analytical approach is thus seen
to be justified for a long constriction. The behavior for shorter
constrictions will be discussed in the next subsection.

An intriguing effect is observed for an elongation $L/L_0 \gtrsim 1.5$.
The force, expected to change sign as soon as the last conductance
channel is closed, \cite{stafford97a} i.e., for $L/L_0 \approx 1.5$,
is found to remain attractive until $L/L_0 \approx 1.6$. For the case
of gold, this would correspond to a stretching of about $5 \AA$ beyond the
point where the conductance becomes vanishingly small. This surprising
effect may actually have been observed.\cite{rubio96}

We have checked whether this persistent force originates from 
our arbitrary choice of the shape of the constriction. 
Therefore, we have determined the shape that minimizes the 
macroscopic part of the free energy\cite{kassubek99} for a fixed 
elongation. With our constraint, this is equivalent to minimizing the 
boundary length of the constriction, subject to the condition of constant
area. 
We find that in this case the effect is even stronger than in other 
geometries. 
Nevertheless, in three dimensional wires, this persistent 
force depends strongly on the shape and sometimes even disappears. 
Further work is needed to clarify the origin of this 
interesting effect. 
It is worthwhile to mention that for this optimized 
shape the conductance steps turn out to be more pronounced and closer 
to experiments than for other shapes. 

\pagebreak
\subsection{Non-adiabatic constrictions}\label{subsec:res-nonadiab}

Our next goal is to go beyond the limitations of the adiabatic 
approximation and to study short constrictions. 
To this end, we consider clean systems with a fixed radius $k_F R=6$,
and vary the initial length $L_0$.
We characterize the geometry of a wire by its opening angle $\theta$,
defined by $\tan\theta=2R/L_0$.

In the adiabatic approximation the wavefunction of Eq.\
(\ref{eq:H}) is factorized, $\psi(x,z)=\phi^z(x)\chi(z)$, and 
terms containing derivatives of $\phi^z(x)$ with respect to $z$ 
are neglected. 
This solution corresponds to a set of decoupled channels. 
In the present context, the approximation is valid if
$\left|\frac{1}{r}\frac{dr}{dz}\right|\ll k$, where $k$ is the
wave vector of the incident electron.
For the case of the conductance, where only states at the Fermi energy are 
involved, this condition is equivalent to $\theta\ll\pi/2$.
Figure \ref{fig:nonadiab}(a) shows the conductance versus elongation 
$L/L_0$ and opening angle $\theta$.
These results for the conductance are consistent with those of
Torres et al.:\cite{torres94}
for an adiabatic wire ($\theta\ll\pi/2$), the conductance
is well quantized in units of $2e^2/h$, with sharp steps between the
plateaus, while for non-adiabatic wires the edges are rounded off and the
plateaus are no longer well defined.
The effects of the geometry on the force are shown in Fig.\
\ref{fig:nonadiab}(b).
One notices that as $\theta$ increases, the average force becomes larger and 
the force oscillations are slightly suppressed.   The overall increase of the 
force is due to the increased effect of surface tension in short constrictions.
In the two-dimensional case, the surface is one-dimensional, and the surface
energy is proportional to the circumference\cite{kassubek99} 
\begin{equation}
\Omega_{\rm surf}=\frac{2\varepsilon_F}{3\lambda_F}
\partial A = \frac{4\varepsilon_F}{3\lambda_F}\int_0^L dz\, 
\left[1+(dr/dz)^2\right]^{1/2}.
\end{equation}
In the WKB approximation, the circumference is approximated
by $\partial A_{\rm WKB}= 2L$, which is a good approximation provided
$(dr/dz)^2\ll 1$, as pointed out in Ref.\ \onlinecite{stafford97a}. However, 
this approximation leads to an underestimate of the surface tension in 
short constrictions, where $|dr/dz|$ is large.  The increased effect of
surface tension in short constrictions with a special wide-narrow-wide
geometry was recently discussed by Kassubek et al.\cite{kassubek99}
As to the force oscillations, we note that they remain well-defined 
provided that one can resolve the conductance plateaus. 
Thus we attribute their decrease with increasing values of $\theta$ 
to enhanced tunneling. 

We remark that wires with larger $\theta$ can 
be stretched more than ones with small $\theta$, that is to say, the cohesive
force remains negative (attractive) further past the point where the 
conductance goes to zero. 
Thus the remanent cohesion is enhanced for non-adiabatic constrictions.

\subsection{Effects of disorder}\label{subsec:res-dis}

We now turn to the study of disorder effects. 
We first calculate the conductance for a single impurity configuration, 
then construct histograms from several hundred samples and 
compare them both with experimental results and previous numerical
and theoretical work.
Finally, we study the effect of disorder on the tensile force.

\subsubsection{Conductance for a single impurity configuration}
\label{subsubsec:res-dis-cond}

\begin{figure}
\begin{center}
\psfig{figure=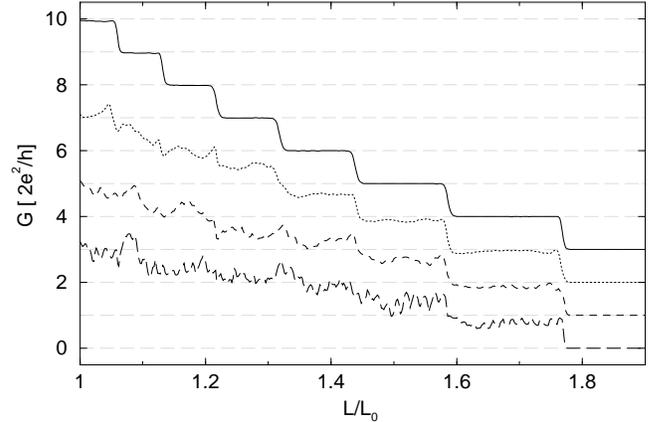,width=8.6cm}
\end{center}
\vspace{-0.5cm}
\caption{Effects of disorder on the conductance of a wire of width
$2 k_F R=24$ and initial length $k_F L_0=60$. 
The same configuration of impurities located both within the constriction 
and in the adjacent regions, as in Fig.\ \ref{fig:geometry}, 
is used for each curve. 
The disordered region is given by $k_F L_{dis}=36$, 
and the impurity concentration is in all cases $\lambda_F^2 n_{dis} = 0.27$. 
The impurity strengths are, from top to bottom,
$k_F^2 \gamma = 0.29, 2.90, 7.26 \text{ and }
21.8\,\varepsilon_F$,
corresponding, respectively, to mean free paths 
$k_F \ell = 7000, 70, 11 \mbox{ and } 1.2$.
Curves are shifted by one unit of conductance.}
\label{fig:disorder}
\end{figure}

We model an impurity at $\vec{r_j}=(x_j,z_j)$ by a local 
potential $V(x,z) = \gamma\delta(x-x_j)\delta(z-z_j)$.
The disorder for a density $n_i$ of impurities 
can be characterized by the mean free path in the Born approximation, 
$\ell =\frac{\hbar^4k_F}{m^2n_{i} \gamma^2}$.
Fig.\ \ref{fig:disorder} shows conductance
curves for a disordered wire with 7 occupied channels
($k_F R=12,\, k_F L_0=60$), with an impurity concentration
$\lambda_F^2 n_{i} = 0.27$ both within the constriction and over a
length $k_F L_{dis} = 36$ on each side of the constriction
(see Fig.\ \ref{fig:geometry}). The disorder strengths $\gamma$
correspond to mean free paths in the range $1.2\leq k_F\ell\leq 7000$.
The spatial distribution of impurities is the same for all curves.

For a very large mean free path $k_F\ell=7000$ (solid line in Fig.
\ref{fig:disorder}), the quantization is almost unaffected by the
disorder. When $\ell$ decreases, higher conductance plateaus are
destroyed. Lower plateaus remain well defined at first, although
they are shifted to lower values (see dotted line in Fig.\
\ref{fig:disorder}). This shift is smallest for the lowest plateaus.
For strong disorder, $k_F\ell \approx 1$, the conductance
quantization is no longer visible except for the very first step. 

This behavior is similar to some experimental results,
\cite{costa-kramer97a,brandbyge97a}
where only a few conductance steps can be observed,
with a shift to conductance values lower than integer
multiples of $2e^2/h$. 
This shift is experimentally observed to
increase for higher conductance values, in agreement with our
numerical results.

\subsubsection{Conductance histograms}\label{subsubsec:res-dis-hist}

\begin{figure}
\begin{center}
\psfig{figure=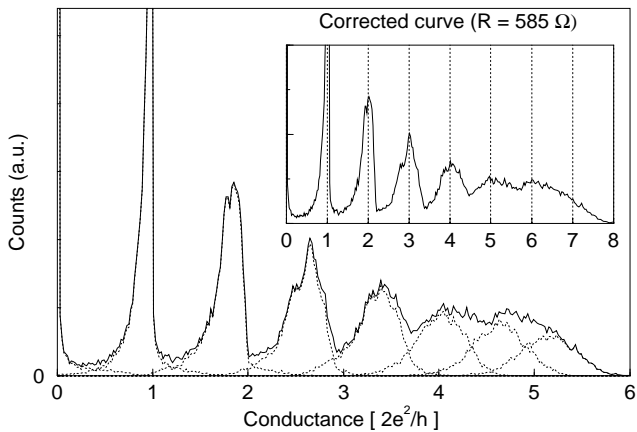,width=8.6cm}
\end{center}
\vspace{-0.5cm}
\caption{Conductance histogram made out of 380 individual conductance
curves.
The width $2R$, initial length $L_0$, length $L_{dis}$ and 
impurity concentration $n_i$ are chosen as in Fig.\ \ref{fig:disorder}. 
The impurity strength is $k_F^2 \gamma=2.9\,\varepsilon_F$ 
(mean free path $k_F\ell=70$). 
The inset shows the same histogram corrected by the calculated sheet
resistance of $R_s\simeq 585\,\Omega$.}
\label{fig:hist}
\end{figure}

As pointed out by several authors,\cite{costa-kramer97c,brandbyge97a}
conductance measurements are reproducible only in very controlled
experiments and hence it is easier to study conductance
quantization statistically in terms of histograms.\cite{olesen94,krans95,costa-kramer97c}
In Fig.\ \ref{fig:hist}, we show a conductance histogram representing 
380 different impurity configurations.  
The same geometry, impurity concentration, and disorder strength as 
for the dotted line of Fig.\ \ref{fig:disorder} have been used, 
corresponding to a mean free path $k_F\ell = 70$. 
Individual peaks (shown as dotted lines in Fig.\ \ref{fig:hist} ) can 
be resolved by constructing histograms from partial conductance curves 
for elongations $L$ in the interval $L_{n+1}<L<L_{n}$, 
where $L_n$ is the midpoint of the $n$-th step of the clean system, 
i.e. $G(L_n)=(2e^2/h) (n-1/2)$. 

The peaks of the histogram are not located at integer
multiples of $2e^2/h$, but are instead shifted to lower values,
as is expected from the single curve of Fig.\ \ref{fig:disorder}.
They can be moved back to quantized values of the conductance,
as for experimental results,\cite{costa-kramer97a}
by subtracting a classical resistance in series with that caused by 
the constriction. 
This additional contribution can be estimated using the Drude formula 
for the conductivity, $\sigma = \frac{n e^2\ell}{mv_F}$, 
where $n$ is the electronic density. 
For the mean free path chosen in the simulations we obtain a sheet 
resistance $R_s\simeq 585\,\Omega$. 
This value of the resistance is consistent with the ones found in
experiments (see for example Refs.\
\onlinecite{brandbyge95,costa-kramer97a,costa-kramer97b,costa-kramer97c}). 
Subtracting it from the total resistance, we indeed find that 
the peaks are shifted to quantized values,
as is shown in the inset of Fig.\ \ref{fig:hist}.

\begin{figure}
\begin{center}
\psfig{figure=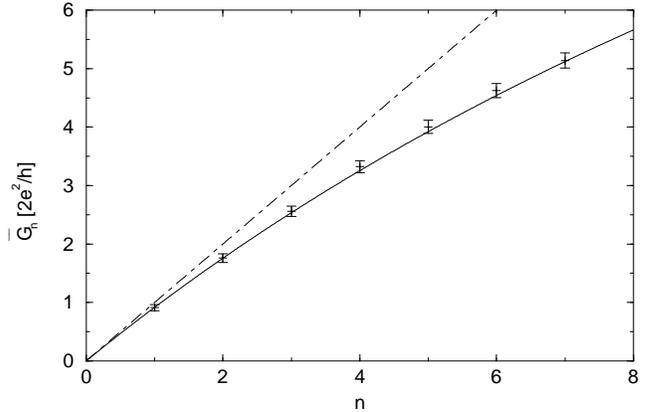,width=8.6cm}
\end{center}
\vspace{-0.5cm}
\caption{Mean values and widths of the peaks of the conductance histogram 
of Fig.\ \ref{fig:hist}. The bars show the numerical results, while the 
solid line gives the mean value as predicted from random matrix theory. 
The dotted line represents the conductance of the clean system and is shown 
for comparison. }
\label{fig:beenakker}
\end{figure}

The resolution of individual peaks allows one to calculate the mean value 
${\bar G_n}$ and root-mean-square width $\Delta G_n$ of the $n$-th peak. 
These quantities have been calculated by Beenakker and Melsen using 
random matrix theory (RMT) for a slightly 
different situation.\cite{beenakker94} 
They have a set-up where {\sl (i)} the disordered and constricted parts are 
spatially separated by scattering-free segments, 
{\sl (ii)} the widening between the constricted and unconstricted parts 
occurs abruptly, and {\sl (iii)} the disorder is varied, while the 
constriction remains fixed. 
In our notation, their results for $n$ open channels are given by 
\begin{eqnarray}
{\bar G_n}^{RMT} &=& \frac{2e^2}{h}\left[\frac{n}{1+\gamma_n}-
\frac{1}{3}\left(\frac{\gamma_n}{1+\gamma_n}\right)^3\right] 
\label{eq:Gbar-RMT}\\
\Delta G_n^{RMT} &=& \frac{2e^2}{h}
\left[\frac{2}{15}\left(1-\frac{1+6\gamma_n}{(1+\gamma_n)^6}\right)
\right]^{1/2}
\label{eq:deltaG-RMT}
\end{eqnarray}
where $\gamma_n=(n+1)\frac{2e^2}{h}R_s$. 
These expressions are valid for $\frac{2e^2}{h}R_s\ll 1$, 
i.e. for a good conductor. 
We can mimic the set-up of Ref.\ \onlinecite{beenakker94} by 
sampling the conductance for elongations corresponding 
to the midpoints of the plateaus of the clean wire, where tunneling 
effects are negligible. 
The width $\Delta G_n^i$ is then exclusively due to the different 
impurity configurations. 
Fig.\ \ref{fig:beenakker} shows our numerical results (bars), 
together with the RMT prediction (solid line), 
calculated from Eq.\ (\ref{eq:Gbar-RMT}) using the same sheet resistance 
as in the simulations. 
The agreement between the mean values is excellent, and the widths, 
calculated from Eq.\ (\ref{eq:deltaG-RMT}), deviate less than one percent 
from the numerical results. 
This remarkable agreement may be due to the fact that the impurity 
concentration is low, so that the constriction and 
the disordered parts are effectively separated by a scattering-free region, 
as assumed in Ref.\ \onlinecite{beenakker94}. 
The hypothesis of abrupt widening between the constricted and unconstricted 
regions appears not to be crucial.

\begin{table}
\caption{Comparison of the different widths of the peaks defined in the text}
\label{table:widths}
\begin{tabular}{cccccc}
Peak \# & $\Delta G_n^{tun}$ & $\Delta G_n^i$ &  $\Delta G_n^{RMT}$ &
$\Delta G_n^{tun}+\Delta G_n^i$ & $\Delta G_n^{tot}$\\
\tableline
1 & 0.050 & 0.102 & 0.105 & 0.152 & 0.157 \\
2 & 0.049 & 0.149 & 0.144 & 0.198 & 0.205 \\
3 & 0.045 & 0.177 & 0.176 & 0.222 & 0.218 \\
4 & 0.040 & 0.203 & 0.203 & 0.243 & 0.249 \\
5 & 0.035 & 0.226 & 0.225 & 0.261 & 0.257 \\
6 & 0.028 & 0.247 & 0.244 & 0.275 & 0.276 
\end{tabular}
\end{table}
\nopagebreak
The histogram of Fig.\ \ref{fig:hist}
also contains a broadening $\Delta G_n^{tun}$ due to tunneling, which 
depends on the shape of the constriction and can be estimated from 
the results for the clean system. The total widths are found to be 
well reproduced by simple sums of impurity and tunneling contributions, 
$\Delta G_n^{tot} \approx \Delta G_n^i +\Delta G_n^{tun}$, as shown 
in Table \ref{table:widths}. 

Finally, we want to emphasize that our result is closer to experiment 
than previous numerical calculations,
\cite{garcia-mochales97} in particular with respect to the peak heights, 
which decrease strongly with increasing conductance. 
We attribute the improved agreement
to the fact that we consider an ensemble of different geometries, with 
fixed Fermi energy, as in the experiments, while in Ref.\ 
\onlinecite{garcia-mochales97} the Fermi energy was varied.  Since these 
systems are nearly ballistic, the ergodic behavior\cite{lee85}
expected in the diffusive regime is violated: an ensemble of different 
geometries with fixed $\epsilon_F$ is not statistically equivalent to 
an ensemble over $\epsilon_F$ with fixed geometry.

\begin{figure}

\begin{center}
\psfig{figure=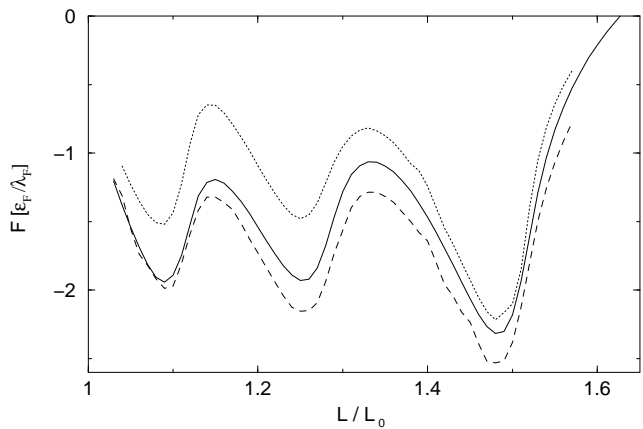,width=8.6cm}
\end{center}

\caption{Impurity effects on the tensile force for a wire
of width $2k_F R=12$, initial length $k_F L_0 = 30$ and a 
disorder characterized by $k_F L_{dis} = 60$ and $k_F\ell=54$ 
(upper and lower curves). 
The middle line represents the force of the corresponding clean system.}
\label{fig:dis-force}

\end{figure}

\subsubsection{Force}\label{subsubsec:res-dis-force}

In this last subsection, we study the effects of disorder on the tensile
force.
In this case, it is no longer optimal to use a delta-function 
impurity potential. 
The force appears to be more sensitive to discretization 
than the conductance. 
In fact, the lattice constant $a$ must be smaller than the range 
of the impurity potential. 
Therefore we use a Gaussian impurity potential with a finite extent $d>a$. 

The wire studied has a geometry given by $k_F R=6$ and $k_F L_0=30$.
The disordered length is $k_F L_{dis}=60$ with an impurity concentration
$\lambda_F^2 n_{i}=0.27$, corresponding to a mean free path $\ell = 54$. 
Results are shown in Fig.\ \ref{fig:dis-force} for two different 
configurations of disorder.
Although the disorder is rather weak, 
it has a strong effect on the tensile force, and can either weaken 
or strengthen the cohesion. 
An intuitive understanding of these opposite behaviors can be obtained by 
considering the change in the density of electrons due to disorder. 
An impurity located at the center of the constriction (c.f. upper 
curve of Fig.\ \ref{fig:dis-force}) would deplete the electron density in the  
constriction, and thus weaken the cohesion. 
Correspondingly, impurities located away from the center of the 
constriction tend to increase the density of electrons in the center, 
thus increasing the cohesion (c.f. lowest curve of 
Fig.\ \ref{fig:dis-force}). 
Further work is needed to understand these effects quantitatively.

\section{Conclusions}\label{sec:conclusions}

In this paper, we have presented an exact numerical solution of the
free-electron model of metallic nanowires,\cite{stafford97a}
using recursive Green's function techniques. 
Our method allows, in principle, to calculate both the conductance 
and the cohesive properties for wires of arbitrary shape, but for 
computational reasons we have limited ourselves to 
two-dimensional wires containing a constriction with a simple geometry. 

The validity of the adiabatic and WKB approximations
used in Ref.\ \onlinecite{stafford97a} was confirmed for long 
constrictions, while for short constrictions the cohesion was
found to be enhanced. 
A persistence of the tensile force after closing of the last conducting 
channel has been observed. It is not yet clear if this is an important 
physical effect or just a curiosity due to the reduced dimensionality. 

We have commonly assumed an ad hoc shape $r(z)$ of the constriction, 
which in general does not represent equilibrium. A more consistent way 
would be to calculate the equilibrium shape 
which minimizes the free energy for a given elongation. 
We have made a first step in 
this direction by minimizing the macroscopic part of the free energy. 
We have found that in this case the wire can be stretched more before 
it breaks. 

Extensive calculations have been made to 
investigate the effects of random impurities.  
Our model gives conductance histograms very similar to those 
observed experimentally. 
The peaks tend to decrease in height and broaden with increasing conductance.
This confirms that disorder in the nanowire is the cause of the
resistance that is usually subtracted from experimental results. 
The positions and widths of the peaks are in excellent agreement 
with calculations based on random matrix theory.\cite{beenakker94} 
We have shown that the widths have two components, 
one due to impurities (perfectly described within random matrix theory), 
the other due to tunneling, which depends on the shape 
of the constriction. 

Our calculations have shown that the effect of disorder is 
particularly strong on cohesion, which is sensitive to the specific 
impurity distribution at the center of the constriction. 
In fact, depending on the presence or absence of an impurity 
close to the center of the constriction, 
the cohesion is either decreased or increased. 
This finding, though qualitatively understandable, 
requires further study. 

Our model may appear to be oversimplified in several respects. 
We have used hard-wall boundary conditions, neglected both the 
electron-electron interactions and the ionic background, and assumed 
a specific geometry upon elongation. 
However, we do not think that these assumptions are crucial for the 
essential effects studied in this paper. 
Other boundary conditions hardly affect the results. 
Coulomb interactions are found to be barely visible in the 
cohesion,\cite{yannouleas98,stafford98} and are believed 
to preserve conductance quantization.\cite{safi97} 
As to the ionic 
background, one has to worry on the one hand, about its effects on 
the electronic structure, and on the other hand, about the kinetics of the 
deformation. The first effect may produce interesting fine structure 
in the conductance,\cite{sanchez-portal97} while the second can 
lead to different paths for elongation and contraction processes,
\cite{landman90} explaining 
the observed hysteresis.\cite{agrait95} 
A hysteresis effect could be easily 
reproduced within our model by choosing different geometries for the 
two processes. 

Our approach, where the emphasis is on conduction electrons while 
the granularity of matter is neglected, is complementary to molecular 
dynamics simulations, where the atomic rearrangements are followed in 
detail, while the effects of conduction electrons are taken into 
account only in an averaged way, e.g. by the embedded atom method.
\cite{landman90} The comparison between our results and experiments 
shows that, except for hysteresis effects, the overall behavior of 
both cohesive and electronic properties of metallic nanoconstrictions 
are well captured by our very simple model. 

\acknowledgments
We acknowledge helpful discussions with F.Kassubek and H. Grabert. 
This work was supported in part by Swiss National Foundation 
PNR 36 "Nanosciences" grant \# 4036-044033. X.Zotos acknowledges support 
by the Swiss National Foundation grant No. 20-49486.96, 
the University of Fribourg and the University of Neuch\^atel.

\end{document}